# Observation of large scale precursor correlations between cosmic rays and earthquakes

P. Homola[1*], V. Marchenko[24], A. Napolitano[17], R. Damian[10], R. Guzik[10], D. Alvarez-Castillo[1], S. Stuglik[1], O. Ruimi[2], O. Skorenok[9], J. Zamora-Saa[3,4], J.M. Vaquero[7], T. Wibig[16], M. Knap[19], K. Dziadkowiec[10], M. Karpiel[6], O. Sushchov[1], J. W. Mietelski[1], K. Gorzkiewicz[1], N. Zabari[5], K. Almeida Cheminant[1], B. Idźkowski[8], T. Bulik[8,11], G. Bhatta[1], N. Budnev[18], R. Kamiński[1], M.V. Medvedev[20,21], K. Kozak[1], O. Bar[12], Ł. Bibrzycki[12], M. Bielewicz[23], M. Frontczak[12], P. Kovács[15], B. Łozowski[22], J. Miszczyk[1], M. Niedźwiecki[25], L. del Peral[14], M. Piekarczyk[12], M. D. Rodriguez Frias[14], K. Rzecki[10], K. Smelcerz[25], T. Sośnicki[10], J. Stasielak[1], A. A. Tursunov[13]

[1]Institute of Nuclear Physics Polish Academy of Sciences; 31-342 Kraków, Poland.

[2]Racah Institute of Physics, Hebrew University of Jerusalem; Jerusalem IL-91904, Israel.

[3]Departamento de Ciencias Físicas, Universidad Andres Bello; Fernández Concha 700, Las Condes, Santiago, Chile.

[4]Millennium Institute for Subatomic Physics at High Energy Frontier - SAPHIR; Fernández Concha 700, Las Condes, Santiago, Chile.

[5]Jerzy Haber Institute of Catalysis and Surface Chemistry Polish Academy of Sciences; 30-239 Kraków, Poland.

[6]Astroparticle Physics Amateur; 34-500 Zakopane, Poland.

[7]Departamento de Física, Centro Universitario de Mérida, Universidad de Extremadura; Avda. Santa Teresa de Jornet 38, E-06800 Mérida (Badajoz), Spain.

[8]Astronomical Observatory University of Warsaw; Al. Ujazdowskie 4, 00-478 Warsaw, Poland.

[9]Taras Shevchenko National University of Kyiv; 01601 Kyiv, Ukraine.

[10]AGH University of Science and Technology, ul. Mickiewicza 30, 30-059 Kraków, Poland.

[11]Astrocent, Nicolaus Copernicus Astronomical Center; Rektorska 4, 00-614, Warsaw, Poland.

[12]Pedagogical University of Krakow; Podchorążych 2, 30-084, Kraków, Poland.

[13]Institute of Physics, Silesian University in Opava; Bezručovo nám. 13, CZ-74601 Opava, Czech Republic.

[14]Space and Astroparticle Group, University of Alcalá; Ctra. Madrid-Barcelona, Km. 33.7, E-28871 Madrid, Spain.

[15]Wigner Research Centre for Physics; Konkoly-Thege Miklós út 29-33., H-1121 Budapest, Hungary.

[16]Department of Theoretical Physics, Faculty of Physics and Applied Informatics, University of Lodz; Pomorska 149/153, PL-90-236 Lodz, Poland.

[17]University of Napoli "Parthenope", Department of Engineering, Centro Direzionale; Isola C4, 80143, Napoli, Italy.

[18]Irkutsk State University, Physical Department; K.Marx str., 1, Irkutsk, 664003, Russia.

[19]Astroparticle Physics Amateur; 58-170 Dobromierz, Poland.

[20]Department of Physics and Astronomy, University of Kansas; Lawrence, KS 66045, USA.

[21]Laboratory for Nuclear Science, Massachusetts Institute of Technology; Cambridge, MA 02139, USA.

[22]Faculty of Natural Sciences, University of Silesia in Katowice; Bankowa 9, 40-007, Katowice, Poland.

[23]National Centre for Nuclear Research (NCBJ); Soltana Str. 7, 05-400 Otwock-Swierk, Poland.

[24]Astronomical Observatory, Jagiellonian University; Orla Str. 171, 30-244 Kraków, Poland.

[25]Department of Computer Science, Cracow University of Technology, 31-155 Kraków, Poland

*Corresponding author. Email: Piotr.Homola@ifj.edu.pl

**Abstract:** The search for correlations between secondary cosmic ray detection rates and seismic effects has long been a subject of investigation motivated by the hope of identifying a new precursor type that could feed a global early warning system against earthquakes. Here we show for the first time that the average variation of the cosmic ray detection rates correlates with the global seismic activity to be observed with a time lag of approximately two weeks, and that the significance of the effect varies with a periodicity resembling the undecenal solar cycle, with a shift in phase of around three years, exceeding $6\,\sigma$ at local maxima. The precursor characteristics of the observed correlations point to a pioneer perspective of an early warning system against earthquakes.

**One-Sentence Summary:** Variations of secondary cosmic ray detection rates are *periodically* correlated with *future* global earthquake magnitude sum.

**Earthquakes induced by the planetary dynamo?**

Despite decades of research, the mechanisms initiating large earthquakes remain enigmatic[1] which leaves room for testing novel ideas. We propose focusing on the effects that might be triggered by reconfiguration of the planetary dynamo whose mechanisms are associated with the physical processes occurring in the very interior of the Earth. Mass movements inside the Earth could lead to earthquakes (EQ), causing temporary changes in both the gravitational and geomagnetic fields simultaneously. If the changes in the latter propagate relatively fast, they can probably be observed on the surface of the planet earlier than the corresponding seismic activity possibly triggered by gravitational changes. A detection of such precursor effects can be possible, for example, by registering changes in the frequency of detection of secondary cosmic radiation (CR), which is very sensitive to geomagnetic conditions. The existing literature documents the efforts towards identifying transient features of cosmic radiation, solar activity, ionospheric conditions, and the geomagnetic field, that could serve as precursors of seismic effects[2–10], however, none of *cosmo-seismic* or *solar-seismic* correlations have been demonstrated on a global scale so far in a statistically convincing and model independent way (i.e. on a discovery level, see e.g. the criticism concerning the total electron content ionospheric monitoring[11]), and in particular no hypotheses concerning an earthquake precursor effect observable in CR data have been verified. Here we report on an observation of the correlations between variation of the average rates of secondary cosmic ray fluxes measured locally and global seismic activity, and we also point to the periodicity of these correlations (or their observability) which corresponds to sunspot number observations back to the 1960s.

The inspiration for the investigation on the possible earthquake precursor effects in cosmic ray data that precedes this article originates in the research undertaken after the devastating M 8.8 earthquake in Chile, in 2010. The most intriguing results concerning only this particular earthquake include ionospheric anomalies above the earthquake region[12], geomagnetic fluctuations at a distant location[5], and unusual variations of secondary cosmic radiation detection rates[13], all preceding the earthquake by different time periods: 15 days, 3 days, and ⅓ day, respectively. The latter result, i.e. the unusual secondary cosmic ray rate , was recorded by the Pierre Auger Observatory (Auger), the largest cosmic ray infrastructure, dedicated mostly to research related to ultra-high energy cosmic rays, but also offering interdisciplinary opportunities such as space weather studies with their scaler data[13, 14]. The Auger site is located in Argentina, ~500 km away from the Chilean earthquake epicenter, thus a good candidate location to probe the possible connection between the secondary cosmic ray fluxes and this particular seismic event. While the Auger studies concerning the big Chilean earthquake were not published, they triggered a longer term interest diffused within the cosmic ray community, resulting in reviving the related research under the scientific agenda of the Cosmic Ray Extremely Distributed Observatory (CREDO)[15] - a recent cosmic ray initiative dedicated mostly to the global search for large scale cosmic ray correlations and associated inter-domain efforts, e.g. those related to the joint research program of the astroparticle physics and geophysics communities[16]. The first extensive *cosmo-seismic* studies of the CREDO programme concerned the public Auger scaler data set, and they were focused on short term scale (up to few days) correlations of secondary cosmic ray detection rates and precursor effect searches using the major (magnitude $\geq 4$) earthquakes with epicenters located at different distances (up to 7,000 km) from the Pierre Auger Observatory. The apparent inconclusiveness of these still-ongoing studies triggered an alternative, novel approach on which we report here: comparing the absolute average variabilities of secondary cosmic radiation to the average global sum of earthquake magnitudes.

Since we consider seismogenic processes occurring very deeply under the Earth's surface it is justified to widen the search for manifestations of *cosmo-seismic* correlations on the surface of the Earth to global phenomena - just because one can attribute no "locality" to deep interior processes. A consequence of this approach is that instead of individual major earthquakes and the corresponding before- and aftershocks, one has to pay attention to the earthquake events occurring globally within a specific time window. Both of these consequences have been adopted in the analysis presented here.

**The cosmo-seismic correlations**

To look for the correlations between the detection rates of secondary cosmic rays and seismic activity we explored the Pierre Auger Observatory scaler data(*13*) compared to selected stations of the Neutron Monitor Database (NMDB)(*17*), and to the earthquake data from the U.S. Geological Survey (USGS) database(*18*). In addition, as a reference for the space weather situation, solar activity data were taken into account, available from the Solar Influences Data analysis Center (SIDC)(*19*). All the data sets used within this study are illustrated in Fig. 1 using a binning relevant for the analysis to be presented subsequently.

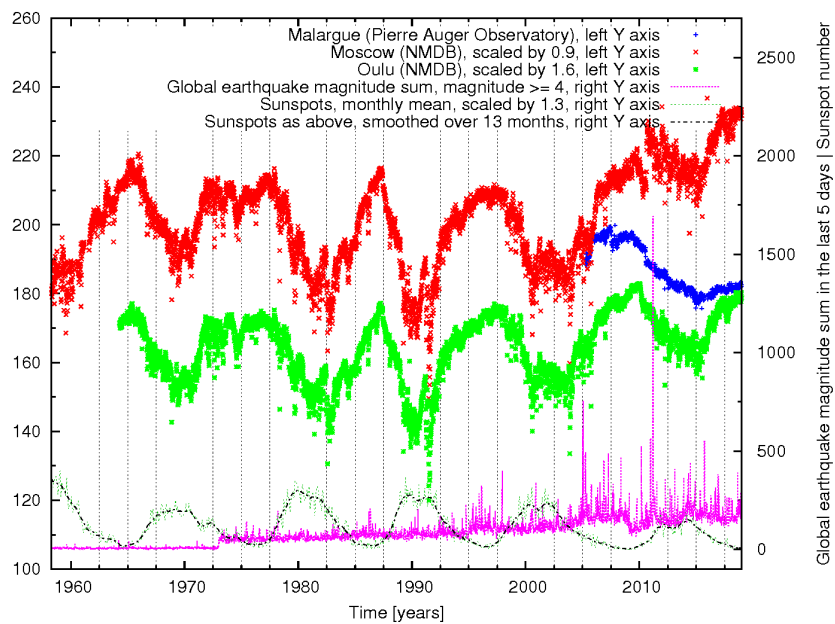

**Fig. 1: The data sets analyzed in this study**. The points in the earthquake and cosmic ray data sets correspond to values averaged over the previous 5 days. Solar activity is visualized with monthly averages of sunspot numbers, and with monthly averages smoothed over the period of 13 months.

The hypothesized complexity of the physical connection between magnetohydrodynamics of the interior of the Earth and the subsequent variations of the secondary cosmic ray detection rates justifies no *a priori* expectations concerning the proportionality between cosmic and seismic data. It is not even clear which kind of cosmic ray response to seismicity should be expected: a specific strength of a transient magnetohydrodynamic instability of the planetary dynamo might result in different seismic effects, depending on the location of the instability with respect to the seismically sensitive regions, and could give a complex picture of the corresponding geomagnetic fluctuations. The subsequent variations of the cosmic ray detection rates might then in principle possess characteristics which are different from those of the corresponding seismic activity, while the two effects might still remain correlated or even causally connected. In particular, neither the direction of changes of the cosmic ray rates nor the direction of their changes can *a priori* be expected to reflect the corresponding behavior of the seismic data. On the other hand, within the planetary dynamo mechanism, one can expect a *change* in the *variations* of the cosmic ray data to be caused by some mass reconfiguration in the Earth's interior. One also considers the inertia of the planetary dynamo system: slow movement of the liquid iron in the Earth interior (reflected in the variation of the cosmic ray rates) might trigger a seismic effect only after some threshold of resistance of the adjoining matter (rocks) is exceeded. It then motivates the search for transient changes of the geomagnetic field and, consequently, of the CR flux before a rapid increase of the global earthquake number. In consequence, an adequate and sufficiently general approach for checking whether there is any correlation between seismic activity and cosmic radiation seems to be a dichotomization of the data(*20*) which would turn the analysis into a simple *yes/no* study by allowing the application of binomial distribution in order to assess the statistical significance of the possible effects. In addition, one is inclined to introduce a time-dependent parameter which could reflect the potential precursor character of the expected correlations. We define the expression

$$c_i(d, m, t_0, t_i, \Delta t, P) \; = \; A_i(d, m, t_0, t_i, \Delta t, P) \, \times \, B_i(d, t_0, t_i, P) \qquad (1)$$

where

$$A_i(d, m, t_0, t_i, \Delta t, P) \; = \; \frac{S_m(d, m, t_0, t_i + \Delta t)}{M(S_m(d, m, t_0, t_i + \Delta t), P)} - 1, \text{ and } B_i(d, t_0, t_i, P) = \frac{|\Delta n_{CR}(t_0, t_{i,i-1})|}{M(|\Delta n_{CR}(t_0, t_{i,i-1})|, P)} - 1$$

where $\Delta n_{CR}(t_0, t_{i,i-1}) = n_{CR}(t_i) - n_{CR}(t_{i-1})$ is the difference in the average cosmic ray detection rates between the two neighbor intervals ending at $t_i$ and $t_{i-1}$, $S_m(d, m, t_0, t_i + \Delta t)$ is the global sum of the earthquake magnitudes larger than or equal to *m* during the corresponding interval ending at $t_i + \Delta t$ with $\Delta t$ being the time shift of the earthquake data set with respect to the cosmic ray data, $M(S_m, P)$ and $M(|\Delta n_{CR}|, P)$ are the medians of the corresponding quantities over the period of length $P$ within which the search for the correlations is being checked, $t_0$ determines the starting time of the period $P$, and $d = t_i - t_{i-1}$ specifies the length of the time interval over which the cosmic ray rates are averaged and the sums of magnitudes $S_m$ are calculated. Then for a given set of free parameters: $P$, $t_0$, $d$, $m$, and $\Delta t$, one defines variables $N_{+/-}$ as sums of positive/negative signs of expression (1) to obtain the binomial probability density function (PDF) describing the probability of getting exactly $k$ positive signs for $n$ intervals of length $d$, over the period $P$:

$$P_{PDF}(N_{+/-} = k) = \left(\frac{n!}{k!(n-k)!}\right) p_{+/-}{}^{k} (1 - p_{+/-})^{n-k} \qquad (2)$$

with $p_{+/-}$ being the probability of a "success": getting the positive/negative sign of expression (1). One expects the following five situations that determine the sign of expression (1):

    I.   $sign(c) = (+) \times (+) > 0,$
   II.   $sign(c) = (-) \times (-) > 0,$
  III.   $sign(c) = (+) \times (-) < 0,$
  IV.   $sign(c) = (-) \times (+) < 0,$
   V.   $sign(c) = 0.$

The situation $V$ can occur if one or more data values are equal to the median value, e.g. in case of odd $n$. If we require that

$$(A_i \neq 0) \; and \; (B_i \neq 0) \; and \; (n_{CR}(t_i) > 0) \; and \; (n_{CR}(t_{i-1}) > 0) \; and \; (S_m(t_i + \Delta t) > 0), \quad (3)$$

and that in addition the numbers of positive and negative values of $A_i$ and $B_i$ are the same, i.e.:

$$n_{A_i>0} = n_{A_i<0} = n_{B_i>0} = n_{B_i<0}, \qquad (4)$$

then the null hypothesis, defined as independence of the two considered data sets containing the EQ and CR data implies that $p_{+/-} = 0.5$, as the situations $I$, $II$, $III$, and $IV$ might occur with the same probabilities of 25%. Thus the probability $P_{PDF}$ from Eq. (2) and the corresponding cumulative distribution function (CDF) $P_{CDF}(N_{+/-} \geq k)$ can serve as measures to test the null hypothesis.

**The precursor effect**

The search for the *cosmo-seismic* correlations reported here was performed in three stages: at first the Auger data was examined to optimize the search strategy with post-trial checks, then the optimum correlation prescription was fine tuned using technically independent data sets from the Moscow and Oulu NMDB stations within the time period corresponding to the Auger data taking time. Finally, we applied the prescription to the earlier periods of time available in case of the Moscow and Oulu data.

An examination of the Auger data gives a stable and significant correlation result already after a coarse variation of the key parameters, with an example local optimum for: $P = 1675 \, days$, $t_0 = 2 \, Apr \, 2014 \, 22{:}07{:}12 \, GMT$ (apart from the large scale time dependence we have also found a sensitivity to small shifts of the data bins in time, of the lengths less than the individual bin size), $d = 5 \, days$, $m = 4.0$, and $\Delta t = 15 \, days$. Out of the $N = P/d = 335$ intervals 294 fulfilled the requirements (3) and (4) to give $N_+ = 202$ and $N_- = 92$ (the sum of $N_+$ and $N_-$ is less than $N$ because of a number of empty intervals in the Pierre Auger Observatory data - such intervals were excluded from the analysis), with the corresponding medians $M(S_{m}, P) = 859.55$ and $M(|\Delta n_{CR}|, P) = 0.48$, with $P_{PDF}(N_+ = 202, \, N = 294) = 3.5 \times 10^{-11}$ and $P_{CDF}(N_+ \geq 202, \, N = 294) = 6.3 \times 10^{-11}$. The latter value corresponds to the significance of more than $6.5 \, \sigma$.

The prescription found for the Auger data was then applied to other cosmic ray data sets, recorded by the aforementioned Moscow and Oulu NMDB stations. In the Moscow case, when applying all the free parameter values exactly as in the prescription, one receives the CDF significance at the level of ~$3.5 \, \sigma$ ($N_+ = 199$, $N_- = 132$), and when we allow a role of the local properties of the Moscow site manifesting in the change of the starting time $t_0$, a scan of this parameter beginning from 30 March 2005 (the data of the first record in the Auger database) with a step of 6 hours (the data bin width in the Moscow database), i.e. checking 13,400 partly overlapping periods, reveals again an effect at the level of ~$6 \, \sigma$ ($P_{CDF} = 4.1 \times 10^{-9}$; $N_+ = 218$, $N_- = 113$) for $t_0 = 14 \, Nov \, 2013 \, 07{:}00{:}00 \, GMT$, i.e. ~4 months earlier and with borders of five-day intervals on a different time of the day compared to the effect observed for the Pierre Auger Observatory site (morning in Moscow vs. evening at the Auger site). Similarly, the Oulu NMDB data reveal a sharp minimum chance probability of ($P_{CDF} = 1.6 \times 10^{-9}$; $N_+ = 220$, $N_- = 112$) at $t_0 = 4 \, Jan \, 2014 \, 23{:}37{:}12$, yet another value of the starting time, more or less in the middle between the values found in the Auger and Moscow data. The comparison between the Auger, Moscow and Oulu results with the corresponding sunspot numbers is presented in Fig. 2.

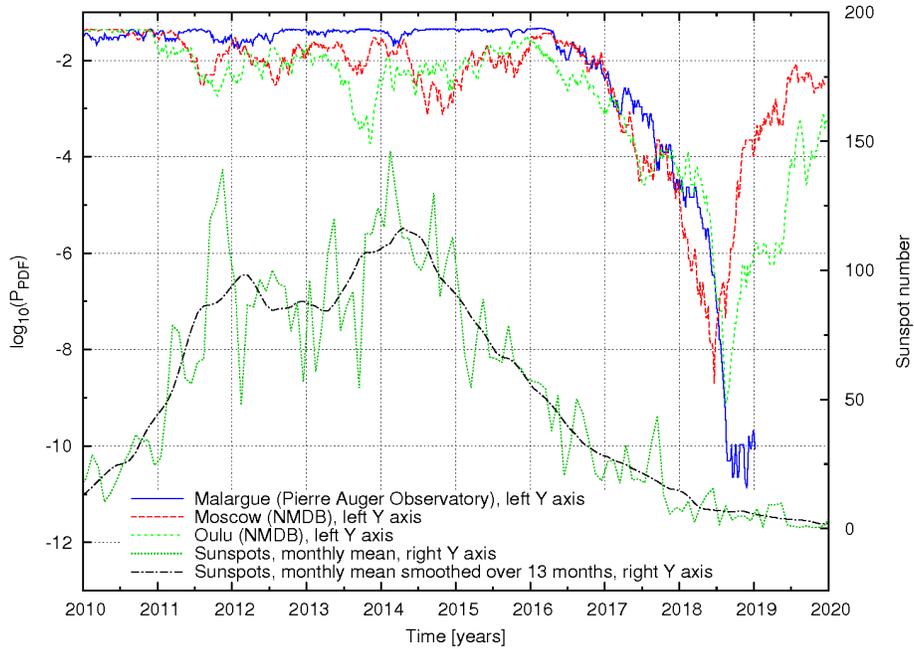

**Fig. 2: ~6 $\sigma$ significance of the effect in three technically independent CR data sets collected by the Moscow and Oulu NMDB stations, and by the Pierre Auger Observatory, compared to sunspot numbers.** Each point illustrates the correlation effect during the last ~4.5 years (335 five-day intervals). All the significance curves were obtained after fine tuning of the parameter $t_0$ performed by applying 20 small shifts in time between 0 and 5 days.

An important feature of the apparent *cosmo-seismic* correlations is a sensitivity of the significance of the effect on the time shift between the CR and EQ data sets. The strongest correlations are found for $\Delta t = \sim 2\ weeks\ (15\ days)$, which points to the precursor character of the CR data behavior with respect to the seismic data changes, as illustrated in Fig. 3 using the Auger data.

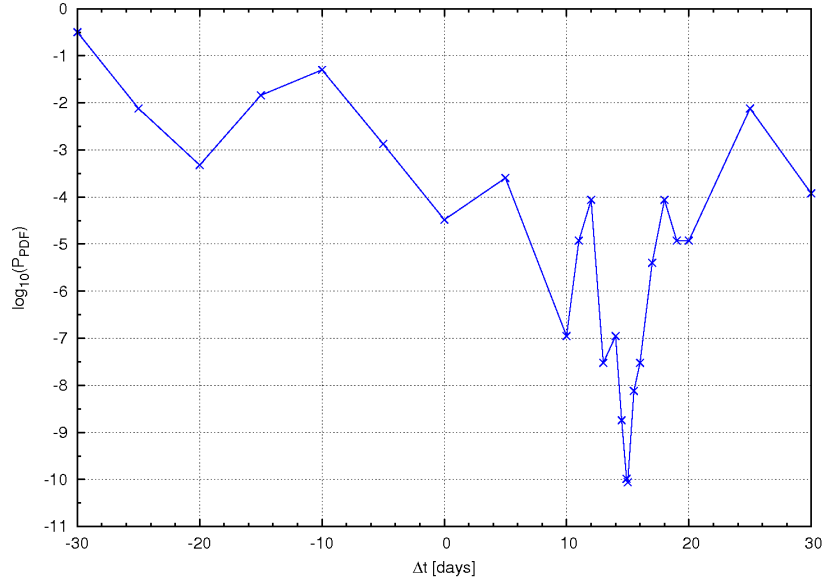

**Fig. 3: The dependence of the significance of the *cosmo-seismic* correlations on the time shift $\Delta t$ of the EQ data with respect to the Auger CR data, for the optimum free parameter set defined in Eq. 1.** The positive or negative values of $\Delta t$ correspond to the situations in which one compares the secondary cosmic ray data in a given time interval to the seismic data recorded in time intervals in the future or in the past, respectively.

While the apparent chance probabilities of the correlation effect are very low in all the three CR data sets, the statistical significance of the result has an uncertainty due to fine tuning of the free parameters needed to find the lowest $P_{PDF/CDF}$, and related to the physical correlations between the CR data sets introduced by the solar activity. However, a simple verification of the significance of the demonstrated relation between the CR and EQ data can be performed by "looking elsewhere", i.e. by considering earlier periods of data taking in the available detectors.

**The statistical significance and the role of the Sun**

The large-scale time dependence and the aforementioned uncertainty of the apparent *cosmo-seismic* correlation seen in Fig. 2 motivates a check with an independent data set extending over another period of time, and also using different time windows. Interestingly, such a study including older data reveals excesses of both $N_+$ and $N_-$ varying regularly over time which justifies using the $P_{PDF}$ (Eq. 2) as an indicator of the binomial distribution anomaly, instead of focusing on an excess of a certain type. As illustrated in Fig. 4, applying smoothing windows of two different lengths (~4.5 and ~9 years) to the Moscow data set indicates a connection with the activity of the Sun: between 1965 and 2015 five distinct and significant minima of $P_{PDF}$ are visible when a wider (~9 years) smoothing window is used, all follow sunspot number maxima after ~3 years.

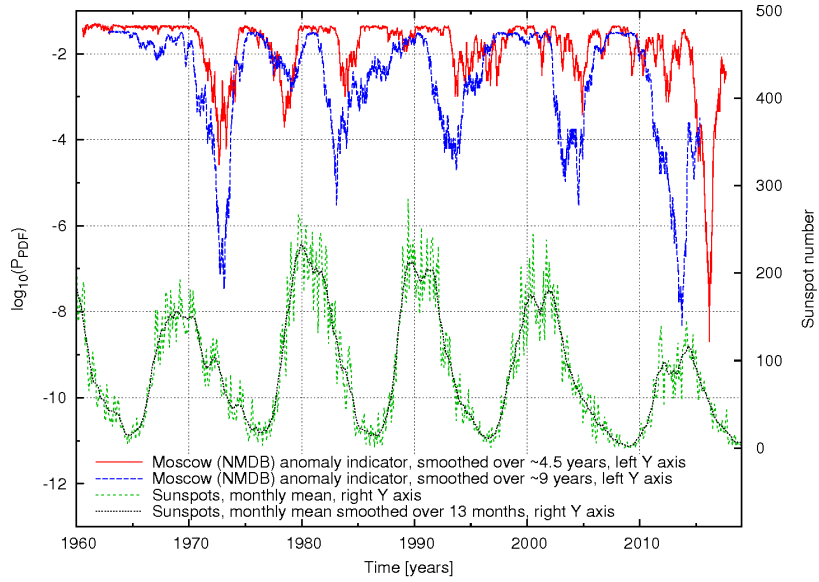

**Fig. 4: The anomaly indicator in the Moscow NMDB data set compared to the sunspot number.** Each point on the correlation significance curves corresponds to the effect found over the smoothing window length of ~4.5 years (1675 days, in red) and ~9 years (3350 days, in blue), with the curve points located at the centers of the windows.

Considering earlier time periods, we scanned over the available $t_0$ range excluding the previously studied period (cf. Fig. 2), and keeping the other free parameters unchanged, i.e. with the same values which gave the results presented in Fig. 2. The procedure gave 3430 new partly overlapping data sets, each of them independent of the excluded "burning sample" data plotted in Fig 2. Using the wider smoothing window of ~9 years results in four new distinct anomalous values of $P_{PDF}$: $3.3 \times 10^{-8}$, $3.0 \times 10^{-6}$, $3.6 \times 10^{-5}$, and $2.9 \times 10^{-6}$ corresponding to $t_0$ values (GMT) of:

1. $12\ Jul\ 1969\ 07{:}37{:}12$,
2. $11\ Jul\ 1978\ 07{:}37{:}12$,
3. $21\ Sep\ 1988\ 07{:}37{:}12$,
4. $23\ Dec\ 1999\ 07{:}37{:}12$,

respectively. For completeness we also list the local minimum $P_{PDF} = 8.1 \times 10^{-9}$, occurring for $t_0 = 23\ Feb\ 2009\ 07{:}37{:}12$. As explained in the previous section, the specific fraction of the five-days period at which we begin the scanning procedure is an effect of the optimization applied already to the "burning sample" of the data, so in the other data set the only factor that penalizes a specific probability value is related to the number of steps in $t_0$ available as new trials. All the four minima occur in non-overlapping periods of 3350 days, so they can be considered as independent events. The overall probability of occurring of all of these minima within the time period checked can then be described by the product of the individual probabilities, with the corresponding penalization factors. To apply these penalization factors to the four minima listed above we use the penalizing factor of 3430 for the first minimum, then for the second minimum the available number of new trials is only 2682 (the scan of the new sets can be continued only beginning from the first $t_0$ value after the previous minimum), for the third minimum it is 1952, and for the fourth one - 1207. Collectively, the probability of an accidental appearance of the four minima to occur during the available data taking time period that precedes the "burning sample" is no greater than

$$3.34 \times 10^{-8} \times 3430 \times 3.0 \times 10^{-6} \times 2682 \times 3.58 \times 10^{-5} \times 1952 \times$$
$$\times 2.92 \times 10^{-6} \times 1207 = 2.3 \times 10^{-10}$$

which corresponds to $\sim 6.3\,\sigma$.

While the recipe we apply does not point to a specific $t_0$ at which an anomalous $P_{PDF}$ should occur, applying a simple scanning rule and the related penalization to compute the chance probability of the *cosmo-seismic* correlation effect to occur many times over decades, confirms that the effect observed is statistically significant. Moreover, the temporal distances between the observed $P_{PDF}$ minima: 10.2, 10.0, 10.2, 11.3, and 9.2 years, as well as the occurrence of the minima ~3 years after the maxima of the solar activity, seem to indicate to a role or even an impact of the Sun that should be studied more deeply in follow up analyses.

**Summary, discussion, and outlook**

We have demonstrated for the first time that the variation of the absolute average detection rates of secondary cosmic radiation correlates with the global seismic situation (sum of the magnitudes of earthquakes with magnitudes greater-than or equal to 4, occurring at all locations) that takes place approximately two weeks later than the relevant cosmic ray data. The size of the shift in time between the cosmic and seismic data sets reveals the precursor character of the correlation effect, coinciding with the time of occurrence of ionospheric anomalies preceding the 2010 8.8-M earthquake in Chile(*12*). The observed correlation effect was validated by independent analyses using cyclostationarity-based methods and randomized data sets (see the Supplementary Information), its significance exceeds 6 $\sigma$, it varies with time with a periodicity resembling the undecenal solar cycle, and it also depends on tiny (less than 5 days), geographically varying shifts of the data bins in time. The latter dependence, although presently not understood, should be investigated interdisciplinarily to search for some lower-level periodicity in the data which might be related e.g. with the rotational period of the Sun, which is approximately 25.6 days at the equator and 33.5 days at the poles, or with tides occurring at maximal strengths twice a month, when the Moon is approximately along the Earth - Sun line.

The $6\sigma$ effect described in this report was found after considering a search for global manifestations of *cosmo-seismic* correlations, without restricting the earthquake data set to the locations of the cosmic ray data used in the analysis. One may suppose that such a result could be the signature of a possible connection between physical mechanisms responsible for changes in the Earth's dynamo and seismic activity. In such a scenario, variations of the geomagnetic field generated by the movements of the liquid core of the Earth could have a direct impact on cosmic-ray detections and would justify the widening of the consideration of *cosmo-seismic* correlations as a global phenomenon observable on the surface of the Earth. However, despite an apparent consistency between the properties of the observed phenomenon, and the geophysical assumptions which motivated the study, at the present stage of the investigation one cannot exclude also non-geophysical interpretations of the periodic *cosmo-seismic* correlations which are demonstrated in this report. For an example, if the solar activity was to induce large scale and energetic transient atmospheric changes which in turn could trigger seismic activity in regions already close to an earthquake due to some other processes, as proposed e.g. in Refs.(*9, 10*) (see also the references therein), the resultant relation between variations of secondary cosmic ray intensities and global seismic activity could look similarly to the effect described here. In any case our observation should be considered as a significant step towards understanding the physics of big earthquakes and to developing an efficient earthquake early warning system.

We expect that the correlations demonstrated here with three arbitrarily chosen independent cosmic ray observation sites should be essentially visible in all the other cosmic ray data sets of comparable quality and volumes, and, possibly, even in smaller sets that extend over a sufficiently long period of time. While precise predictions of seismic activity currently seem unachievable, the fine structure of the observed dependencies, including site-to-site and technique-to-technique differences, creates a perspective of the application of cutting-edge data processing and analysis techniques, including the latest achievements in artificial intelligence and big data, to assess the future earthquake risk at least globally, in a continuous way, and broadcast the information widely, leaving precaution-related decisions to the most exposed governments, organizations, or even individuals. With this starting point concerning the early warning system against earthquakes, the further accumulation of secondary cosmic ray data, together with other *inward* multimessenger channels of physics information, and with continuously improved modeling and methodology of the analyses, the precision of the warnings will only increase and save lives of many throughout the world, wherever the seismic activity is an everyday threat.

We expect that the apparent similarity of the periodical changes of the *cosmo-seismic* correlation effect to the undecenal solar cycle will be a starting point for a new kind of interdisciplinary analyses concerning the yet unconfirmed though possible physical connections (e.g. of magnetic or gravitational origin) between the Sun and the Earth. The subsequent studies should also include, in particular, investigations on shorter time scale precursors, correlations with other known earthquake precursors or precursor candidates (e.g. radon emission and/or particle densities in the ionosphere), and potential connections with the planned technological efforts (e.g. using tiltmeters to obtain the information on gravitational changes that precede seismic effects).

The character and the scope of the potential impact of this study, but also its ultimate relevance, warrant efforts for spreading the presented results as widely as possible so that a collective and well coordinated interdisciplinary research dedicated to an earthquake early warning system can be pursued efficiently.

**Acknowledgments:**

We acknowledge the leading role in the CREDO Collaboration and the commitments to this research made by the Institute of Nuclear Physics, Polish Academy of Science. We acknowledge the Pierre Auger Observatory for providing the scaler data(*14*), the NMDB database (www.nmdb.eu), founded under the European Union's FP7 programme (contract no. 213007), and the PIs of individual neutron monitors at Moscow (Pushkov Institute of Terrestrial Magnetism, Ionosphere and radio wave propagation IZMIRAN, Troitsk, Russia)(*21*) and Oulu (Sodankyla Geophysical Observatory of the University of Oulu, Finland)(*22, 23*) for providing data. This research has been supported in part by the PLGrid Infrastructure - we warmly thank the staff at ACC Cyfronet AGH-UST for their always helpful supercomputing support. We thank Roger Clay, Dariusz Góra, Bohdan Hnatyk, Michał Ostrowski, Leszek Roszkowski, and Henryk Wilczyński for critical reading of the manuscript, and for their useful remarks and valuable discussions.

**Funding:**

Visegrad Fund, Grant No. 21920298 (PH, AT)

National Science Center, Grant No. 2016/22/E/ST9/00061 (VM)

National Science Center, Grant No. 2018/29/B/ST2/02576 (RK, ŁB)

ANID-Millennium Science Initiative Program - ICN2019 044 (JZS)

**Author contributions:**

Conceptualization: PH, VM, AN, DAC, JZS, JWM, NB

Methodology: PH, VM, AN

Resources: PH, MK, BI

Investigation: PH, VM, AN, RD, RG, SS, OR, OSk, KD, MK, OSu, KG

Visualization: PH, SS, MK

Project administration: PH

Supervision: PH, JWM, DAC

Writing – original draft: PH, VM, AN, JZS

Writing – review & editing: PH, DAC, SS, JZS, JMV, TW, NZ, KAC, BI, TB, GB, NB, RK, MVM, KK, OB, ŁB, MB, MDRF, MF, PK, BŁ, JM, MN, LP, MP, KR, KS, TS, JS, AAT

**Competing interests:** Authors declare that they have no competing interests.

**Data and materials availability:** All data are available in public databases cited in the main text.

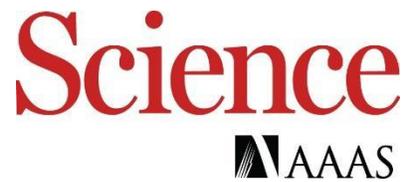

# Supplementary Materials for

## Observation of large scale precursor correlations between cosmic rays and earthquakes

P. Homola, V. Marchenko, A. Napolitano, R. Damian, R. Guzik, D. Alvarez-Castillo, S. Stuglik, O. Ruimi, O. Skorenok, J. Zamora-Saa, J.M. Vaquero, T. Wibig, M. Knap, K. Dziadkowiec, M. Karpiel, O. Sushchov, J. W. Mietelski, K. Gorzkiewicz, N. Zabari, K. Almeida Cheminant, B. Idźkowski, T. Bulik, G. Bhatta, N. Budnev, R. Kamiński, M.V. Medvedev, K. Kozak, O. Bar, Ł. Bibrzycki, M. Bielewicz, M. Frontczak, P. Kovács, B. Łozowski, J. Miszczyk, M. Niedźwiecki, L. del Peral, M. Piekarczyk, M. D. Rodriguez Frias, K. Rzecki, K. Smelcerz, T. Sośnicki, J. Stasielak, A. A. Tursunov

Correspondence to: Piotr.Homola@ifj.edu.pl

**This PDF file includes:**

Materials and Methods
Figs. S1 to S4

**Materials and Methods**

Materials

The main feature of the analysis method proposed in this article is related to putting a focus on a possible long term periodicity of the correlations between locally observed secondary cosmic radiation rates and seismic effects considered globally.

The available Pierre Auger Observatory scaler data used for the study presented here include the records taken between 2005 and 2018[1] which contain the secondary cosmic ray hit rates corrected for the pressure conditions and averaged every 15 minutes over the active stations of the Observatory (up to ~1600 water-Cherenkov detectors, each of ~$10\ m^2$ active surface, located in the province of Mendoza, Argentina). As partly seen in Figure 1 of the main article, the general trends in the Auger data are similar to the reference cosmic ray data from the selected stations of the NMDB network, here the Moscow and Oulu sites, both chosen for this study for their long-lasting and stable performance. The relevance of the two databases (Auger and NMDB) for space weather related studies has been demonstrated in the literature cited in the main article and can be also seen in Fig. 1, mostly by comparing the CR data to the solar activity data, but also by locating the downward outlier drops/points corresponding to the Forbush decreases. These are caused by coronal mass ejections, which cause transient increases of the strength of the geomagnetic field and thus result in the reduction of the CR detection rates. The Moscow and Oulu data, both available as detection rates averaged over intervals of 6 hours, extend back to the late 1950's. As visible in Fig. 1, on a longer term scale both Moscow and Oulu rates behave very similarly, revealing a clear periodicity and anticorrelation with the sunspot number. Although the Auger records span a much shorter period of time compared to Moscow and Oulu, only slightly longer than just one solar cycle, their relevance is easily justified by the overwhelmingly large number of detectors spread over a considerably larger surface compared to point-site and single detectors of the NMDB sites. Moreover, an additional complementarity between the Auger and the chosen NMDB sites is reflected in significantly different locations (southern and northern hemispheres, respectively) and detection techniques (muon/electron water-Cherenkov detectors vs. neutron monitors). One finds several data gaps in the Auger data set, which introduces an additional uncertainty about the final result, although on a conservative side. The discontinuities of the Auger data set are handled with care: when the cosmic ray data is missing, we disregard the EQ data within the corresponding interval. Also, the EQ and CR medians are computed only for CR-EQ interval pairs for which all the data ingredients have non-zero values (cf. condition (3) in the main text).

As explained in the main article, the seismic data considered within this study were processed as global sums of the magnitudes of the earthquakes that occurred within a specific time window, instead of focusing on dates of individual earthquakes only. As shown in Fig. 1, introducing an example window of 5 days and the minimum earthquake magnitude of 4, demonstrates a feature-rich distribution on small, medium and large scales. Except for a clear, long-term deficit before the year 1973 which suggests inefficiency of earthquake reporting, one observes a general increasing trend beginning around 1973 and lasting for decades since then, several medium features lasting for a couple of years, and numerous short-term activity periods, lasting from a few days to several months. Such a richness of the scales and features additionally justifies a focus on a parameter which describes the Earth seismic activity in a global way.

Methods

The results obtained in the article were validated using cyclostationarity-based methods and with randomized data sets.

*Validation with cyclostationarity-based methods*

Aimed at corroborating the results presented in the main article, statistical dependence or correlation are analyzed by studying the joint cyclostationarity properties of pairs of the following time series: cosmic ray detection rate, earthquake sum magnitude, and sunspots.

The cyclostationary model is appropriate when signals are created by the interaction of periodic and random phenomena. In such a case, the signal itself is not periodic, but the periodicity is hidden and is present in its statistical functions (*24*, *25 chaps. 1-2*). Data originated by geophysical phenomena exhibit cyclostationarity due to, for example, Earth revolution and rotation (*25 sec. 10.9*, *26*). Data originated by astrophysical phenomena exhibit cyclostationarity due to revolution and rotation of stars and planets and periodicities in Sun and star pulsation and activity (*25 sec. 10.9*, *27*).

Cyclostationary feature measurements show that pairs of the considered time series can be suitably modeled as jointly cyclostationary. The estimated Fourier coefficients of the periodically time-varying joint distribution function or of the joint cross-correlation function are different from zero in correspondence of frequencies related to characteristic periods of the time series.

**Cyclostationarity Analysis**

In the considered problem, the signals $y_1(n)$ and $y_2(n)$ are single time series, that is, for each of them an ensemble of realizations, namely a stochastic process, does not exist. In such a case, the statistical characterization is more suitably made in the functional of fraction-of-time (FOT) approach (*24*, *25 chap. 2*, *28*). In the FOT approach, starting from a single time series, all familiar probabilistic parameters such as mean, autocorrelation, distribution, moments, and cumulants, are constructed starting from the unique available time series.

In the FOT approach, the joint cumulative distribution function (CDF) of $y_1(n + m)$ and $y_2(n)$ is defined as

$$F_{y_1 y_2}\left(n, m; \xi_1 \xi_2\right) \triangleq P\left[y_1(n+m) \leq \xi_1, y_2(n) \leq \xi_2\right] = E^{\{\alpha\}}\left\{u\left(\xi_1 - y_1(n+m)\right) u\left(\xi_2 - y_2(n)\right)\right\}$$

where $P[\cdot]$ denotes FOT probability (*24, 25*), $u(\xi) = 1$ for $\xi \geq 0$ and $u(\xi) = 0$ for $\xi < 0$, and $E^{\{\alpha\}}\{\cdot\}$ is the almost-periodic component extraction operator, that is, the operator that extracts all the finite-strength additive sine-wave components of its argument. It is the expectation operator in the FOT approach.

All the results can be interpreted in the classical stochastic approach by interpreting $P[\cdot]$ as classical probability and $E^{\{\alpha\}}\{\cdot\}$ as the ensemble average $E\{\cdot\}$, provided that appropriate ergodicity conditions (called cycloergodicity conditions) are satisfied by the stochastic processes (*24, 25 chap. 5*).

The function $F_{y_1 y_2}\left(n, m; \xi_1 \xi_2\right)$ is almost-periodic in $n$ by construction. For jointly stationary time-series it does not depend on $n$. We have (*25 sec. 2.3.1.5*)

$$F_{y_1 y_2}\left(n, m; \xi_1 \xi_2\right) = \sum_{\alpha \in \Gamma_2} F_{y_1 y_2}^\alpha\left(m; \xi_1, \xi_2\right) e^{j2\pi\alpha n}$$

where $\Gamma_2$ is a countable set of possibly incommensurate cycle frequencies $\alpha \in [-1/2, 1/2)$ and

$$F_{y_1 y_2}^\alpha\left(m; \xi_1, \xi_2\right) \triangleq \lim_{N \to \infty} \frac{1}{2N+1} \sum_{n=-N}^{N} u\left(\xi_1 - y_1(n+m)\right) u\left(\xi_2 - y_2(n)\right) e^{-j2\pi\alpha n}$$

are the Fourier coefficients which are referred to as cyclic joint CDFs. The function $F_{y_1 y_2}^\alpha\left(m; \xi_1, \xi_2\right)$ is not identically zero for $\alpha \in \Gamma_2$. For jointly stationary time-series, $F_{y_1 y_2}^\alpha\left(m; \xi_1, \xi_2\right)$ is non zero only for $\alpha = 0$.

The cross-correlation function of $y_1(n)$ and $y_2(n)$ is given by

$$E^{\{\alpha\}}\left\{y_1(n+m)\, y_2(n)\right\} = \int_{R^2} \xi_1\, \xi_2\, dF_{y_1 y_2}\left(n, m; \xi_1 \xi_2\right) = \sum_{\alpha \in \Gamma_2} \int_{R^2} \xi_1\, \xi_2\, dF_{y_1 y_2}^\alpha\left(m; \xi_1, \xi_2\right) e^{j2\pi\alpha n} =$$

$$= \sum_{\alpha \in A_2} R_{y_1 y_2}^\alpha(m)\, e^{j2\pi\alpha n}$$

where $A_2 \subseteq \Gamma_2$ is a countable set of possibly incommensurate cycle frequencies $\alpha \in [-1/2, 1/2)$ and

$$R_{y_1 y_2}^{\alpha}(m) \triangleq \lim_{N \to \infty} \frac{1}{2N+1} \sum_{n=-N}^{N} y_1(n+m) \, y_2(n) e^{-j2\pi\alpha n} = \int_{R^2} \xi_1 \, \xi_2 \, dF_{y_1 y_2}^{\alpha}(m; \xi_1, \xi_2)$$

are the Fourier coefficients which are referred to as cyclic cross-correlation functions. The function $R_{y_1 y_2}^{\alpha}(m)$ is not identically zero for $\alpha \in A_2$.

## Measurement Results

The Fourier coefficients $F_{y_1 y_2}^{\alpha}(m; \xi_1, \xi_2)$ and $R_{y_1 y_2}^{\alpha}(m)$ of the periodically time-varying cross statistical functions of pairs of time series $y_1(n)$ and $y_2(n)$ are estimated over the available finite observation intervals ($N$ finite in the above expressions) (*25 chap. 5, 29 chap. 2*). The energies of the estimates, for each $\alpha$, are the sum over lag parameter $m$ of the squared magnitudes of the estimates.

Missing values in the data files are reconstructed by linear interpolation. Then, time series are dichotomized. In the joint CDF, the temporal median values $\xi_1$ and $\xi_2$ of $y_1(n)$ and $y_2(n)$ are considered.

The analyzed time series have been obtained with sampling periods of 5 days or 1 month. Each sample of a time series sampled with sampling period 1 month is replicated 6 times in order to obtain a new time series with sampling period 5 days (1 month = 30 days = 6 × 5 days). Thus, all processed time series have the same sampling period of 5 days. The sampling period is denoted by $T_s$ and the sampling frequency by $f_s = 1/T_s$. When time series have different lengths, the shortest is zero-filled.

## Statistical Dependence Between Cosmic Rays and Earthquakes (Experiment 1)

In the first experiment,

- Time series $y_1(n)$ is the average variation of the cosmic ray detection rate taken from Moscow (NMDB), (original sampling period = 5 days);

- Time series $y_2(n)$ is $log_{10}$ of earthquake (EQ) sum magnitude taken from Moscow (NMDB), (original sampling period = 5 days).

Results for the estimated cyclic joint CDF $F_{y_1 y_2}^{\alpha}(m; \xi_1, \xi_2)$ are reported in Fig. S1.

Significant cyclostationary features are found at cycle frequencies $\alpha = \pm \, 0.001288 \, f_s$. The cycle frequency $\alpha_0 = 0.001288 \, f_s$ corresponds to the period $T_0 = 776.02 \, T_s = 3880.1$ days = 10.63 years.

## Cyclic Cross-Correlation Between Cosmic Rays and Sunspots (Experiment 2)

In the second experiment,

- Time series $y_1(n)$ is the average variation of the cosmic ray detection rate taken from Moscow (NMDB), (original sampling period = 5 days);

- Time series $y_2(n)$ is the Sunspot monthly mean (original sampling period = 1 month).

Results for the estimated cyclic cross-correlation function $R^{\alpha}_{y_1 y_2}(m)$ are reported in Fig. S2.

Significant cyclostationary features are found at cycle frequency $\alpha = \pm \ 0.00130 \, f_s$. A more detailed analysis shows features at cycle frequencies $\alpha = \pm \ 0.001101 \, f_s$ and $\alpha = \pm \ 0.001466 \, f_s$. These cycle frequencies merge into $\alpha = \pm \ 0.00130 \, f_s$ in the shown graphs. The cycle frequency $\alpha_0 = 0.00130 \, f_s$ corresponds to the period $T_0 = 769.2 \, T_s = 3846.2$ days = 10.53 years.

## Cyclic Cross-Correlation Between Earthquakes and Sunspots (Experiment 3)

In the third experiment,

- Time series $y_1(n)$ is $log_{10}$ of earthquake (EQ) sum magnitude taken from Moscow (NMDB), (original sampling period = 5 days);

- Time series $y_2(n)$ is the Sunspot monthly mean (original sampling period = 1 month);

Results for the estimated cyclic cross-correlation function $R^{\alpha}_{y_1 y_2}(m)$ are reported in Fig. S3.

Significant cyclostationary features are found at cycle frequency $\alpha = \pm \ 0.001279 \, f_s$. The cycle frequency $\alpha_0 = 0.001279 \, f_s$ corresponds to the period $T_0 = 781.5 \, T_s = 3907.5$ days = 10.71 years.

## Conclusion

Measurements of joint cyclic statistical functions show the existence of statistical dependence or correlation between the pairs of analyzed signals. In particular, the dark lines in Fig. S2 left and Fig. S3 left is the evidence of joint cyclostationarity of the analyzed pairs of signals. The cycle frequencies obtained from the analysis correspond to periods of almost 11 years.

*Validation with randomized data sets*

In order to validate the significance of cosmo-seismic correlations that was obtained after applying "local optima" of the parameters to each data set, we have conducted additional validation using the randomized data sets. For such a purpose the cosmic ray (CR) and earthquake (EQ) data were downloaded and binned independently according to the prescription from the main article.

Following the approach described in the article we introduce the additional parameter $X_{CR/EQ}$ for CR and EQ data in order to characterize their simultaneous behavior. Namely, using the variables defined in equation (1) we can write

$$X_{EQ} = A_i(d, m, t_0, t_i, \Delta t, P) + 1 \ , \ X_{CR} = B_i(d, t_0, t_i, P) + 1 \ .$$

In order to study the simultaneous variations of EQ and CR data above their median values, we construct new parameter $X'_{CR/EQ}$ for CR and EQ data by assigning $X'$ to "+1" if its value at a given time step is above its median value ($X_{CR/EQ} > 1$), and to "−1" if the value is below the median value ($X_{CR/EQ} < 1$)

$$X'_{CR/EQ} = \begin{cases} +1, if \ X_{CR/EQ} > 1 \\ -1, if \ X_{CR/EQ} < 1 \end{cases}$$

The variable $N_+$ that was defined in the article and shows the sum of positive signs of expression (1) can be written in terms of $X'_{CR/EQ}$ as

$$N_+ (lag) = \frac{1}{2} \Sigma \left| X'_{CR} + X'_{EQ} \right|$$

where the sum of arrays $X'_{CR}$ and $X'_{EQ}$ is calculated element-by-element and then the sum of its absolute values is calculated for various lags. The non-zero values of this parameter show how many cases we have the situation when CR and EQ behave in a similar way, i.e. both simultaneously go below or above their median values.

We have estimated the uncertainties by random shuffling the original EQ and CR time series, where we have performed N = $10^7$ random realizations and have calculated the correspondent percentiles and moments of obtained distributions of $N_+^{rand}$. The results are shown on Fig. S4. From obtained results we conclude that for each original data set the value of $N_+$ is larger than the highest $N_+$ value in $10^7$ randomized data sets, which is consistent with the main result reported in the article.

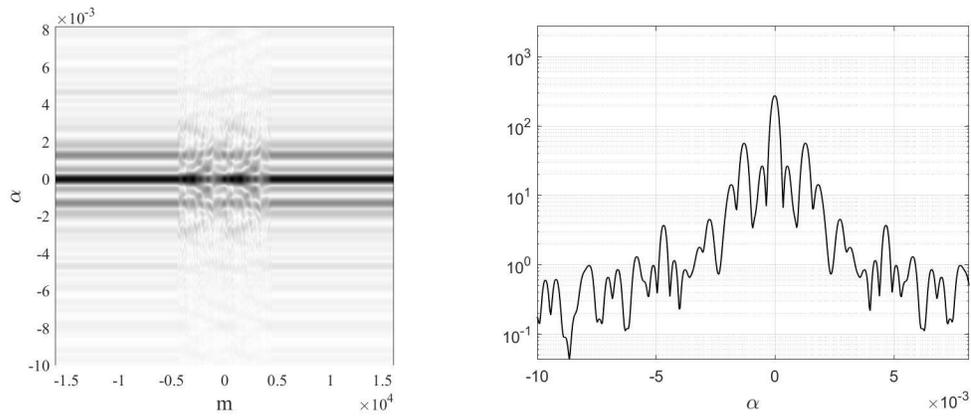

**Fig. S1. Statistical Dependence Between Cosmic Rays and Earthquakes.** (Left) magnitude of the estimated cyclic joint CDF $F_{y_1 y_2}^{\alpha}\left(m; \xi_1, \xi_2\right)$ as a function of $\alpha$ and $m$ (2-dimensional grayscale elevation map); (Right) energy of the estimated joint CDF as a function of $\alpha$.

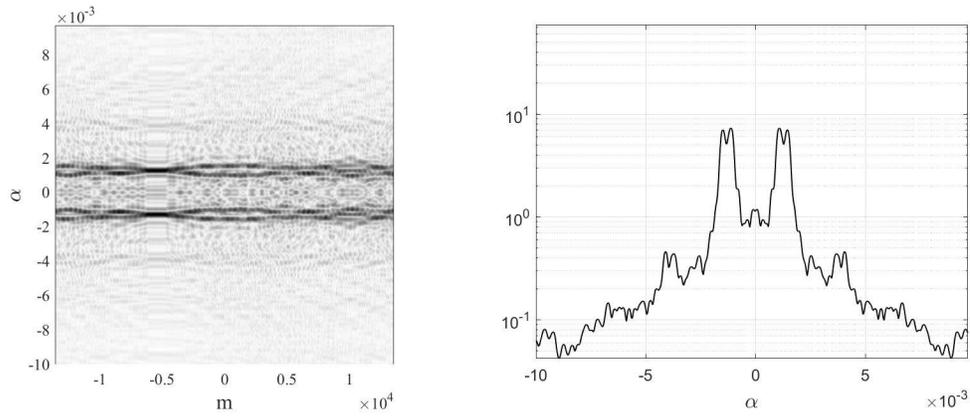

**Fig. S2. Cyclic Cross-Correlation Between Cosmic Rays and Sunspots.** (Left) magnitude of the estimated cyclic cross-correlation function $R_{y_1 y_2}^{\alpha}(m)$ as a function of $\alpha$ and $m$ (2-dimensional grayscale elevation map); (Right) energy of the estimated cyclic cross-correlation as a function of $\alpha$.

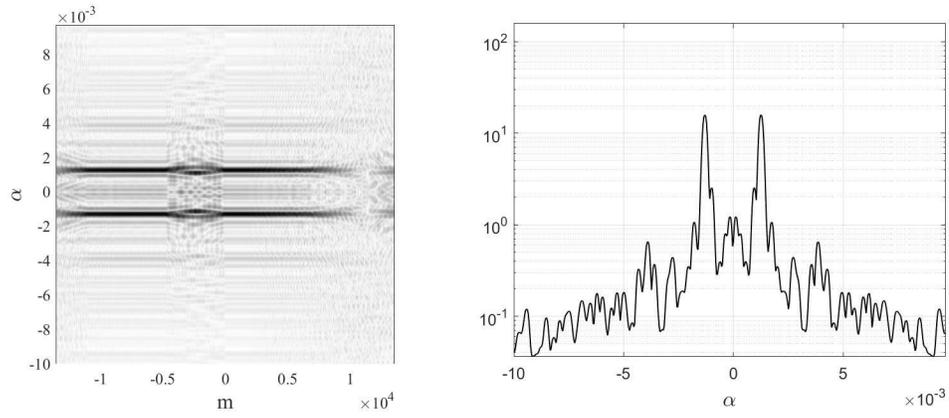

**Fig. S3. Cyclic Cross-Correlation Between Earthquakes and Sunspots.** (Left) magnitude of the estimated cyclic cross-correlation function $R_{y_1 y_2}^{\alpha}(m)$ as a function of $\alpha$ and $m$ (2-dimensional grayscale elevation map); (Right) energy of the estimated cyclic cross-correlation as a function of $\alpha$.

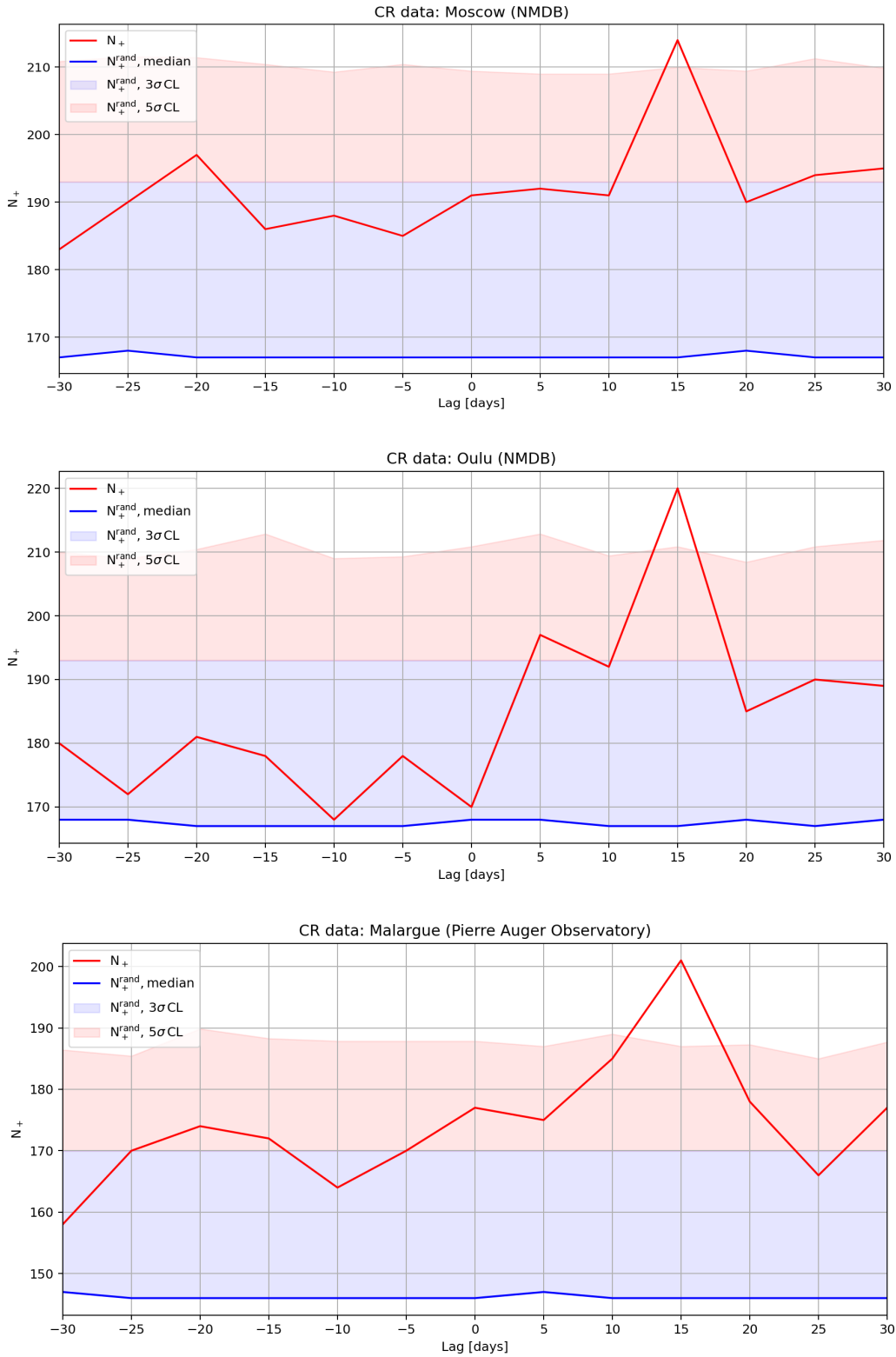

**Fig. S4. The parameter $N_+$ for different time lags for various cosmic ray data.** The median level and $3\sigma$, $5\sigma$ confidence levels from random simulations are shown.